\documentclass[reprint,amsmath,amssymb,aps,prl,superscriptaddress]{revtex4-1}

\usepackage{graphicx}

\begin{document}

\title{Enhancement of Giant Rashba Splitting in BiTeI under Asymmetric Interlayer Interaction}
\author{Taesu Park}
\affiliation{Department of Chemistry, Pohang University of Science and Technology, Pohang 37673, Korea}
\author{Jisook Hong}
\affiliation{Department of Chemistry, Pohang University of Science and Technology, Pohang 37673, Korea}
\author{Ji Hoon Shim}
\email[]{jhshim@postech.ac.kr}
\affiliation{Department of Chemistry, Pohang University of Science and Technology, Pohang 37673, Korea}
\affiliation{Department of Physics and Division of Advanced Nuclear Engineering, Pohang University of Science and Technology, Pohang 37673, Korea}

\begin{abstract}
We carry out density functional theory calculation to enhance the Rashba spin splitting (RSS) of BiTeI by modifying the interlayer interaction. It is shown that RSS increases as the Te layer approaches to adjacent Bi layer or the I layer recedes from the Bi layer. Our results indicate that the RSS can be sensitively increased by introducing a vacancy on the Te site to make effective Bi-Te distance shorter. It is also found that the difference of Te \textit{p} orbital character between two spin-split bands increases when the RSS is developed along crystal momentum, which supports asymmetric interlayer interaction in the spin-split bands. Our work suggests that the modification of interlayer interaction is an effective approach in the modeling of the RSS in BiTeI and other layered materials.
\end{abstract}

\maketitle

\section{Introduction}

Generating spin currents by controlling the spin of materials is one of the goal in spintronics since it is potentially applicable in spin transistor, spin diode, and other electronic devices. \cite{fabi07,manc15,wolf01}. The field of spin-orbitronics has been emerging in the sense that nonmagnetic materials could have spin-dependent properties induced by spin-orbit coupling (SOC). Peculiarly, Rashba spin splitting (RSS) is a phenomenon of lifted spin degeneracy in electronic energy bands due to broken inversion symmetry \cite{rash59,bych84}. Each spin-split state shows locking of spin momentum perpendicular to the crystal momentum \cite{bych84}, which becomes important in manipulating the spin polarization in the viewpoint of engineering.

One of the issue in RSS is the enhancement of Rashba parameters. Among them, the Rashba energy $E_R$ is the energy difference between band extrema and band crossing points in the spin-split bands, and larger $E_R$ has an advantage of stabilizing polarized spin states for spin isolation \cite{liu15}. Bulk BiTeI in our interest has one of the largest spin splitting with $E_R$ around 0.1 eV \cite{ishi11}. In this work, we find that changing the interlayer interaction in BiTeI makes sensitive variations of RSS along with keeping track of the proposed origins of Rashba splitting.

Previously, the origin of RSS has been well studied based on a conventional free-electron model proposed by E. I. Rashba and a localized orbital model under electrostatic field. The former model shows that the spin of an electron couples to the effective magnetic field by its motion in the presence of perpendicular electric field \cite{bych84}. However, the scale of splitting energy in Au(111) surface \cite{lash96} and the relevance with strength of SOC \cite{rein01,pete00} were not explained in this model. To deal with those problems, another model was proposed that the orbital angular momentum (OAM) affects Rashba splitting \cite{park11,park15,hong15}. The local OAM states in crystal momentum induce an asymmetric charge distribution, which couples with surface potential. However, regarding the driving force of giant RSS as an electric field is unrealistic since the required electric field to achieve giant RSS is too colossal \cite{hong}.

So far, there have been a lot of discussions on the origin of the giant RSS in BiTeI. BiTeI has strong SOC and would have large internal potential gradient \cite{ishi11,sunk17}. Large potential gradient from polar field has been proposed to be induced by ionic bond, and this concept has been applied in layered materials \cite{liu13, liu15}. From the viewpoint of microscopic insight, in contrast, it has been found that BiTeI has an orbital-dependent spin texture \cite{zhu13,bawd15} or an anomalous interaction between Bi and I \cite{fu13}. By $\vec{k}\cdot\vec{p}$ perturbation formalism \cite{bahr11}, atomic orbital momentum in potential gradient could be a key for giant RSS, which leaves the possibility of spin-dependent orbital hybridization. For that reason, atomic configuration is important in BiTeI and previous DFT calculation of BiTeI with different atomic coordinate was conducted yielding different spin splitting size \cite{wang13,bahr11}. The calculation implies that the distance between Bi and Te could be related to the spin splitting size \cite{wang13}, so it is normally understood that the RSS in BiTeI is originated by the bulk configuration \cite{ishi11,bahr11}. Therefore, it is meaningful to change and track the atomic interaction of spin splitting in bulk BiTeI. Here, we virtually modified the position of Te and I atomic layers to see the change of RSS and our calculated results are linked to recent multiorbital concept \cite{hong,liu16}. The change of orbital characters along crystal momentum is inspected and the important factor related to RSS is asymmetric interlayer hopping near Bi layer.

\section{Calculation Details}

The band structures of several sets of BiTeI are calculated by using a full-potential linearized augmented plane wave method as implemented in WIEN2K \cite{blah01}. Exchange-correlation potential is chosen to the generalized gradient approximation of Perdew-Berke-Ernzerhof (PBE-GGA) \cite{perd96}. It is known that the space group of BiTeI is $P3m1$ \cite{shev95}, and the lattice constants are given as $a$ = 4.3392 \AA, $c$ = 6.854 \AA. Fractional \textit{z} coordinates of each atom in BiTeI are referred from structural optimization by Bahramy's paper \cite{bahr11}, so the coordinates of Bi, Te and I atoms are (0, 0, 0), (2/3, 1/3, 0.7482), and (1/3, 2/3, 0.3076), respectively. In this work, we consider the shift of Te and I layers by introducing variables $\Delta z_I$ and $\Delta z_{Te}$, and the shift produces new coordinates of (0, 0, 0), (2/3, 1/3, 0.7482+$\Delta z_{Te}$), and (1/3, 2/3, 0.3076+$\Delta z_I$) with same lattice constants. The core separation energy cut-off is set to $-$6.0 Ry and the muffin tin radii ($R_{MT}$) for all atoms are set to 2.5 Bohr radius. All the calculations are conducted considering SOC, and the maximum modulus of reciprocal vector $K_{max}$ is chosen to satisfy $R_{MT} K_{max}$ = 7.0. A \textit{k}-mesh in the Brillouin zone is set to $9\times9\times5$.

\begin{figure}
\includegraphics[width=8.5cm]{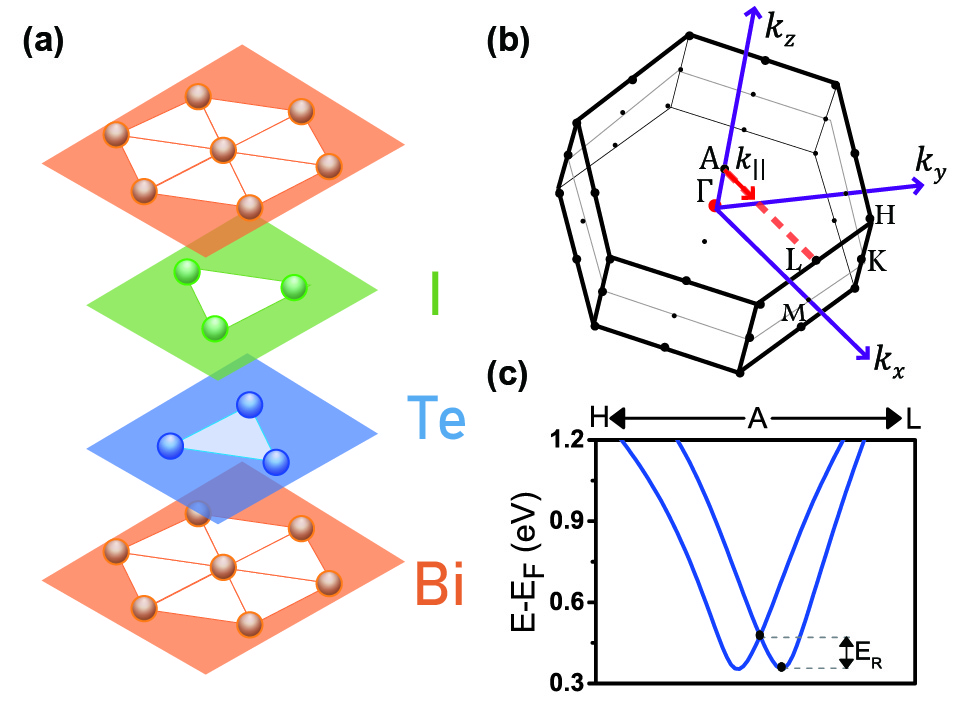}
\caption{ 
(Color online) Structure of BiTeI and its Rashba splitting. (a) Atomic layer configuration of BiTeI showing trilayer configuration. (b) Brillouin zone of BiTeI. $k_{||}$  is defined by the deviation from A point ($k_z$ $=$ $\pi$/$c$) along L point shown by the red arrow. (c) The lowest conduction band of BiTeI along H-A-L direction calculated by DFT calculation. There is Rashba splitting near A point with Rashba energy $E_R$ of 0.118 eV. 
}
\label{p1}
\end{figure}

In the analysis of orbital characters, the linear combination of atomic orbitals (LCAO) coefficients at the spin-split conduction band minimum (CBM) are calculated by using OpenMX code \cite{ozak03,ozak04,ozak05}. Same crystal informations and exchange-correlation potential to WIEN2K calculation are used. SOC is also considered, and the \textit{k}-mesh is set to 6 $\times$ 6 $\times$ 4. Pseudo-atomic basis functions and norm-conserving pseudopotentials are used in OpenMX library. In the calculation, we used \textit{s}$^2$\textit{p}$^2$\textit{d}$^2$ pseudo-atomic basis. We confirm that the band structures from both OpenMX and WIEN2K calculations are the same.

\begin{figure}
\includegraphics{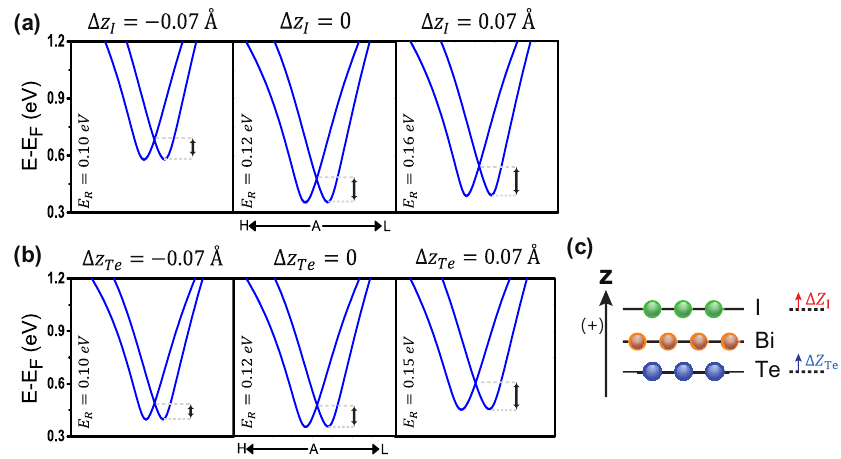}
\caption{
(Color online) The change of band structures at the spin-split CBM for (a) I-shifted, (b) Te-shifted BiTeI. The atomic positions of others are not changed, only \textit{z} coordinate of I (Te) is changed to $-$0.07 \AA , 0.00 \AA , and 0.07 \AA. $E_R$ is increased as I and Te layer move upward. (c) Non-optimized BiTeI with modifiable variables; $\Delta z_I$, $\Delta z_{Te}$.
}
\label{p2}
\end{figure}

\section{Results and Discussions}

The crystal structure of trigonal non-centrosymmetric BiTeI is described in Fig. \ref{p1}(a). It is composed of triply layered unit cell, and its stacking type is ABC stacking, which shows the absence of inversion symmetry. The band structure in Fig. \ref{p1}(c) shows that Rashba splitting of BiTeI is observed in CBM near A point with $E_R \sim$ 0.1 eV, which is consistent with ARPES result \cite{ishi11}. This structure is often considered as semi-ionic model with (BiTe)$^+$ layer, where Bi and Te interact with covalent bonding, and neighboring I$^-$ layer \cite{shev95}. In the vicinity of Fermi level, Bi 6\textit{p} state has dominant contribution to the conduction band and 5\textit{p} states of Te and I are dominant in the valence band. Also, there can be a topological insulating phase under pressure due to narrow band gap of $E_g$ = 0.38 eV \cite{bahr12,xi13,chen13}.

It has been reported that the size of spin splitting is changed by interatomic distance in layered structures \cite{hong, gier10}. In a similar manner, we expect that the change of interlayer distance in bulk BiTeI is crucial to the RSS, so we performed DFT calculations of layer-shifted BiTeI to examine its effect on the electronic band structures and RSS strength. The position of Bi is left to intact, and we consider two variables $\Delta z_I$ and $\Delta z_{Te}$ which are the shifts of atomic layer along \textit{z} direction as indicated in Fig. \ref{p2}(c) within about five percent of \textit{c} lattice parameter. The band structures at the spin-split CBM for I-shift and Te-shift are presented in Figs. \ref{p2}(a) and \ref{p2}(b). The strength of RSS increases linearly as I layer gets far away from Bi layer (Fig. \ref{p2}(a)). On the other hand, Te-shifted case (Fig. \ref{p2}(b)) shows opposite tendency, the strength of RSS increases as Te layer moves toward Bi layer. As interlayer distance is directly linked to the orbital interaction between layers and Bi \textit{p} orbital character is dominant at CBM, it is important to know the interaction of Bi layer with neighboring atomic layers, which will be explained in the later part.

\begin{figure}
\includegraphics[width=8.5cm]{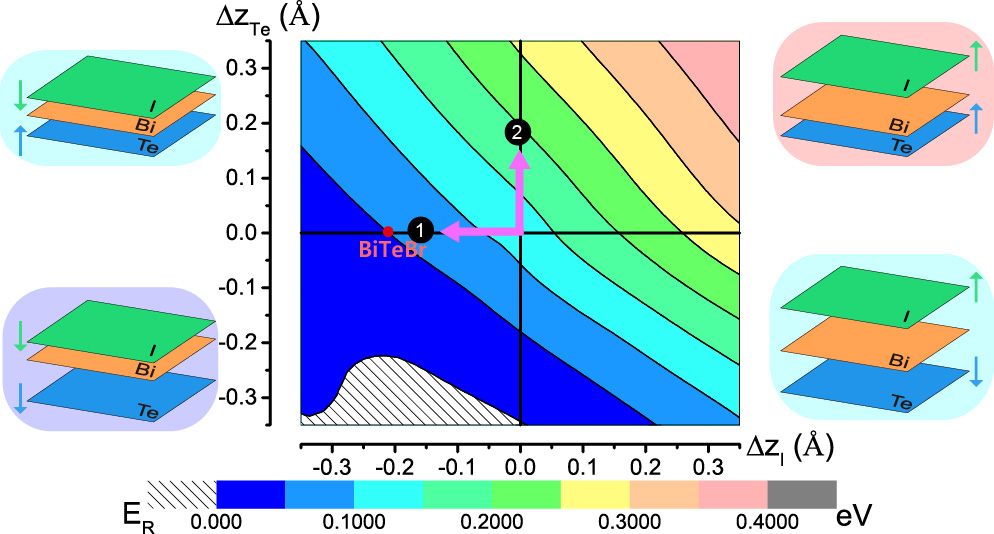}
\caption{
(Color online) The contour plot of $E_R$ in terms of $\Delta z_I$ and $\Delta z_{Te}$. The colormap ranges from red (high $E_R$ up to about 0.4 eV) to blue (small $E_R$) and there is a slash-lined region where $E_R =0$. For each quadrant, the atomic layer shifts are schematized outside the contour with a background colored by corresponding contour level of $E_R$.
}
\label{p3}
\end{figure}

We also consider the shifts of Te and I layers simultaneously, and the contour plot in Fig. \ref{p3} shows tendencies of the RSS strength expressed by $E_R$ with respect to $\Delta z_I$ and $\Delta z_{Te}$. It indicates that the atomic variations with same signs of $\Delta z_I$ and $\Delta z_{Te}$ (Quadrant 1 or 3 in Fig. \ref{p3}) have more sensitive change of the spin splitting than those with opposite signs (Quadrant 2 or 4). Most of previous studies tried to tune the RSS of BiTeI by hydrostatic strain \cite{fu13} or pressure \cite{bahr12}, which correspond to the quadrant 2 or 4 where the change of RSS is not sensitive. However, the case for making the sensitive change of RSS is the shift of only Bi layer or the shift of Te and I layer in the same direction according to Fig. \ref{p3}. Also, it implies that the RSS of BiTeI can be strengthened up to 0.4 eV if the structure could be severely distorted with positive signs of $\Delta z_I$ and $\Delta z_{Te}$. When two variables have both negative signs, the RSS would even disappear as indicated by the slash-lined region in quadrant 3 of Fig. \ref{p3}.

This contour plot is well consistent with the size of spin splitting for non-optimized BiTeI obtained by various structural parameters studied before. The band structures of BiTeI using atomic coordinates from X-ray data and different optimization don't show RSS at CBM \cite{wang13,bahr11}. Te layer in those structures is much further from adjacent Bi layer compared to the optimized structure. Their corresponding $\Delta z_{Te}$ values are $-$0.32 $\sim$ $-$0.38 \AA\ with negative $\Delta z_I$, and they are located near the slash-lined region in contour where there is no spin splitting. In other instances, RSS strength ($E_R$) of BiTeI under hydrostatic strain along in-plane direction changes within the order of 0.01 eV \cite{fu13}. This change is insensitive because hydrostatic strain corresponds to the direction along quadrant 2 or 4 in Fig. \ref{p3}, as in-plane strain also can be considered as expansion or contraction along stacking direction by structural optimization.

Meanwhile, BiTeBr and BiTeCl have smaller spin splitting than BiTeI \cite{erem12}, and Br substituted BiTeI$_{1-x}$Br$_x$ shows reduced spin splitting \cite{wu16} due to the small RSS in BiTeBr compared to BiTeI. It can be seen that $E_R$ decreases as the size of halogen gets smaller, in other words, electronegativity of halogen increases. The interlayer distance between Bi and \textit{X} in BiTe\textit{X} from the reported structural parameters decreases as \textit{X} goes up in the periodic table; 2.1083 \AA\ (BiTeI), 1.8968 \AA\ (BiTeBr), 1.6786 \AA\ (BiTeCl) \cite{zhu13,bahr11,sans16}. This trend is reasonable in the way that smaller halogen makes the distance between Bi and \textit{X} reduced, and it results in the weakening of spin splitting indicated by the arrow along the direction 1 in the contour. If we consider only $\Delta z_I$ assuming interlayer distance between Bi and Te is not varying compared to that between Bi and \textit{X}, this plot can roughly predict the Rashba splitting of BiTe\textit{X} \cite{erem12}. In Fig. \ref{p3}, the corresponding $\Delta z_I$ for BiTeBr is shown by a red dot with $E_R$ of roughly 50 meV, which is similar to 60 meV calculated for BiTeBr \cite{erem12}. In reality, the estimation of RSS size may show some discrepancy because the unit cell size or other distance parameters could be affected for different halogen BiTe\textit{X} and spin-orbit coupling strength for substituted halogen atom is weaker than iodine atom.  $\Delta z_I$ for BiTeCl is $-$0.42 \AA\ which is beyond the contour, and its $E_R$ is expected to be much smaller than the calculated $E_R$, about 30 meV \cite{zhu13}.

Similarly, the RSS in BiTeI can be increased by shifting Te layer upward shown by the arrow along the direction 2 in Fig. \ref{p3}, which reduces the distance between Bi and Te. Due to that, it would be possible to achieve larger splitting by introducing a vacancy on the Te site. Doping of a small atom on the Te site can be expected to achieve larger splitting, so we performed the evaluation of $E_R$ for the doped case by band structure calculation. Fully-doped BiSeI and BiSI from the BiTeI structure show 0.111 eV and 0.114 eV of $E_R$ respectively and 25\% doped BiTeI shows RSS similar to undoped one, whose $E_R$ is 0.117 eV (or 0.119 eV) for Se-doped (or S-doped) BiTeI. Effective Bi-chalcogen interlayer distance decreases as chalcogen atom gets smaller, which provides the evidence that decreased Bi-Te interlayer distance and weakened spin-orbit coupling strength due to dopants of smaller atom counterbalance the spin splitting.

\begin{figure}
\includegraphics[width=8.5cm]{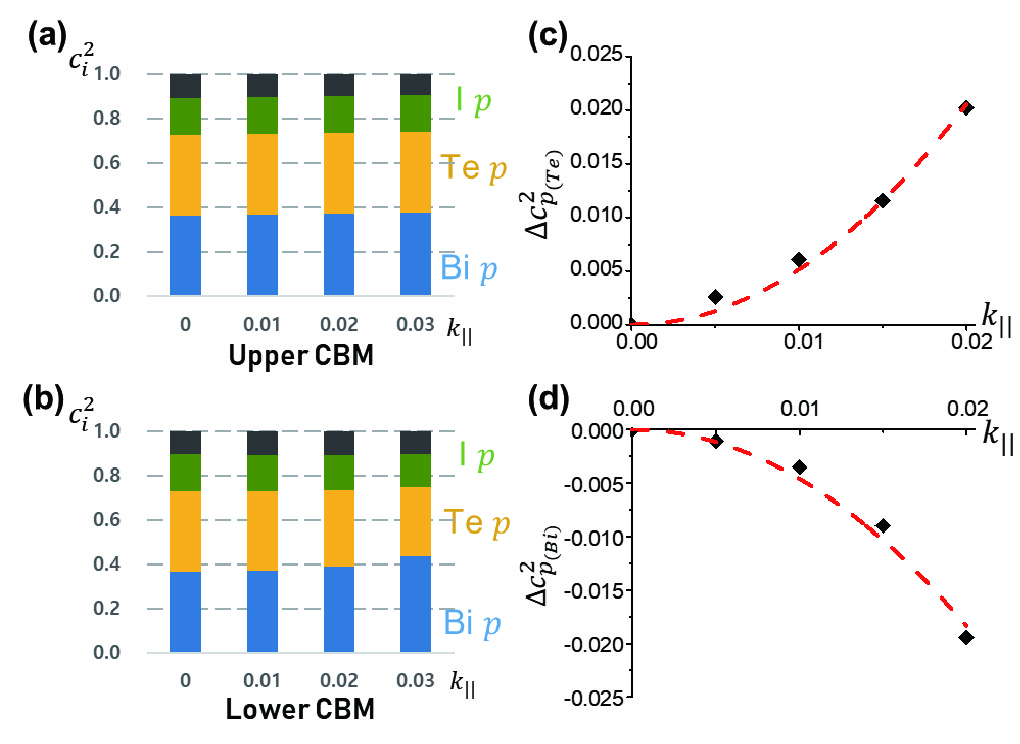}
\caption{
(Color online) Normalized orbital characters of optimized BiTeI at (a) upper band of CBM and (b) lower band of CBM.  Orbital characters of Bi \textit{p}, Te \textit{p}, and I \textit{p} are colored as blue, yellow, and green, the summation of other \textit{s} or \textit{d} orbital characters is expressed as an achromatic color at the top since \textit{p} orbital character is dominant at CBM. Orbital character differences are obtained by subtracting the orbital characters at lower CBM from those at upper CBM, which are plotted for (c) Te \textit{p} orbital and (d) Bi \textit{p} orbital.
}
\label{p4}
\end{figure}

According to the simple tendencies of the contour plot, the sensitive change of RSS in layer-modified BiTeI and scale of $E_R$ imply that asymmetric interlayer atomic interaction is significantly important since the distance between Bi and I ($d_{Bi-I}$) or between Bi and Te ($d_{Bi-Te}$) could be crucial in RSS \cite{fu13,wang13}. Even though each interatomic interaction and spin degree of freedom are entangled complicatedly in BiTeI, we mention that the hopping between Bi and Te is related to RSS by exploring orbital characters in spin-split conduction bands. To analyze interlayer hopping and possible interlayer interaction in this system, the orbital characters at two spin-split CBM are investigated by collecting the orbital characters of optimized BiTeI for given $k_{||}$ indicated in Fig. \ref{p1}(b). They are shown in Figs. \ref{p4}(a) and \ref{p4}(b). First, it can be easily seen that \textit{p} orbitals are dominant and \textit{s} and \textit{d} orbitals are negligibly small, thus it would be sufficient to consider only \textit{p} orbital characters in this system. Next, the orbital characters at upper CBM don't vary significantly under $k_{||}$, but Te \textit{p} character at lower CBM decreases when $k_{||}$ increases. As a result, Te \textit{p} character difference between two spin-split CBM increases along $k_{||}$ as shown in Fig. \ref{p4}(c) whereas Bi \textit{p} character difference shown in Fig. \ref{p4}(d) decreases complementary to Te \textit{p} character, and they make hopping difference between Bi and Te for two spin-split CBM.

\begin{figure}
\includegraphics[width=8.5cm]{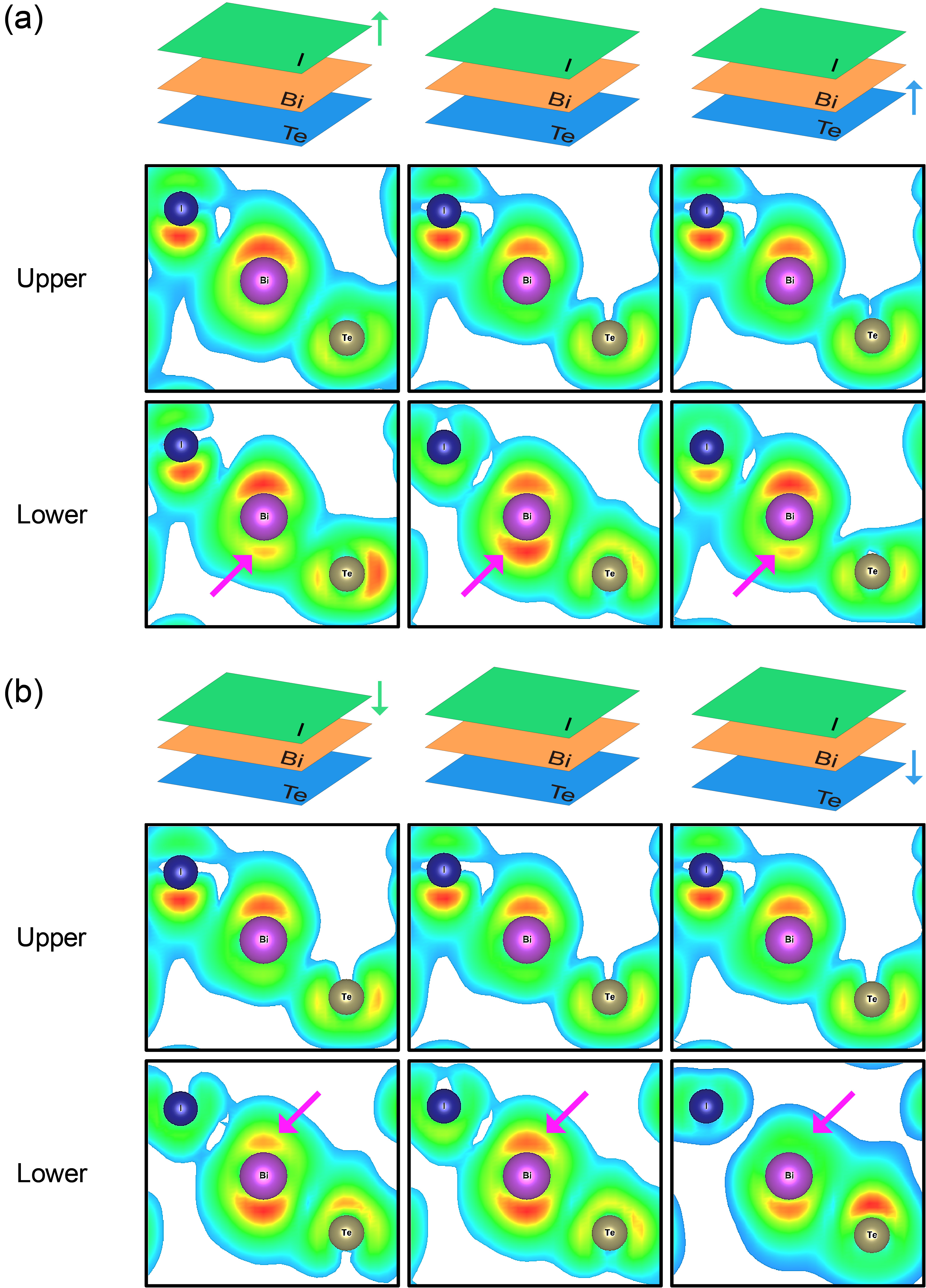}
\caption{
(Color online) Partial charge density of spin-split CBM in layer-shifted BiTeI at $k_{||}$ = 0.04 in unit of 2$\pi$/a. (a) The case when Rashba splitting is increased. (b) The case when Rashba splitting is reduced. For all cases, displacement of atomic layer from the optimized position is 0.07 \AA\, and corresponding band structures can be found in Fig. \ref{p2}. Middle figure shows the charge density for optimized BiTeI.
}
\label{p5}
\end{figure}

Since different charge distribution between spin-split bands can induce the giant RSS in layered system \cite{hong}, we directly analyzed the change of wave function for spin-split CBM by inspecting partial charge density ($|\psi_{CBM}|^2$) in real space when Te or I layer is shifted. As shown in Fig. \ref{p5}, there is a correlation between partial charge density near the Bi atom and RSS variation from the optimized BiTeI. Charge density profile for the upper band shows insignificant change whenever any layer is shifted, which implies negligible variation of orbital character. However, the charge density in lower CBM shows the different tendency for RSS indicated by arrows near Bi atom. If the spin splitting is increased by displacing Te and I layer upward, the density between Bi and Te atom becomes sparser (Fig. \ref{p5}(a)). When RSS is reduced from the optimized BiTeI, upper lobe of Bi atom shows decreased charge density (Fig. \ref{p5}(b)). This contrasting behavior means the difference of orbital interaction, which is dependent on the changing size of Rashba splitting. Therefore, the RSS size modified by changing the interlayer interaction would be related to the asymmetric interatomic hopping with neighboring layers.

\section{Conclusions}

BiTeI has giant RSS and it has been proposed that atomic configuration is important for this phenomenon. To investigate the spin splitting for the enhancement of RSS, we directly modified the atomic interaction by shifting the atomic layer positions of Te and I along the \textit{z} direction in DFT calculation. The result shows that the RSS size increases as Te layer approaches to Bi layer and as I layer gets away from Bi layer. Also the change of RSS is sensitive when $\Delta z_I$ and $\Delta z_{Te}$ move in the same direction along the stacking axis, so it can be enhanced significantly in principle.
From the result, we expect that the different orbital character in each spin states which is originated from atomic interaction could explain this splitting. Also, we find that orbital character difference between Bi and Te in each spin-split bands is related to the RSS strength. As RSS is developed under crystal momentum far from the degenerate point, we see that Te \textit{p} orbital contributions are different between the spin-split bands indicating the asymmetric interatomic hopping. This approach would be helpful for linking to currently studying multiorbital model for RSS \cite{bawd15,liu16}, estimating the dynamical dependence of RSS such as lattice vibration \cite{mons17}, and finding better spin-polarized materials. Shifting the atomic layer artificially in bulk may be deadly realizable since it might destruct the stable crystal configuration, but a variation in atomic scale such as a vacancy would be possible to enhance the RSS.

\end{document}